\documentclass[pre,twocolumn,floatfix, superscriptaddress,a4paper,nofootinbib]{revtex4-1}
\usepackage{amsmath,bm,graphicx}
\usepackage{amssymb}
\usepackage{amsfonts}
\usepackage{epstopdf}
\usepackage{mathrsfs}
\usepackage{color}
\usepackage{latexsym}
\usepackage{hyperref}
\usepackage{subfigure}
\usepackage{bm}
\usepackage{natbib}
\usepackage[normalem]{ulem}

\begin{document}

\title{Phoretic colloids close to and  trapped at fluid interfaces}

\author{Paolo Malgaretti}
\email[Corresponding Author: ]{p.malgaretti@fz-juelich.de }
\affiliation{Helmholtz Institute Erlangen-N\"urnberg for Renewable Energy (IEK-11), Forschungszentrum J\"ulich, Cauer Stra{\ss}e 1,
91058 Erlangen, Germany}

\author{Jens Harting}
\affiliation{Helmholtz Institute Erlangen-N\"urnberg for Renewable Energy (IEK-11), Forschungszentrum J\"ulich, Cauer Stra{\ss}e 1,
91058 Erlangen, Germany}
\affiliation{Department of Chemical and Biological Engineering and Department of Physics, Friedrich-Alexander-Universit\"at Erlangen-N\"urnberg, F\"{u}rther Stra{\ss}e 248, 90429 N\"{u}rnberg, Germany
}

\begin{abstract}
The active motion of phoretic colloids leads them to accumulate at boundaries and interfaces. Such an excess accumulation, with respect to their passive counterparts, makes the dynamics of phoretic  colloids particularly sensitive to the presence of boundaries and pave new routes to externally control their single particle as well as collective behavior.
Here we review some recent theoretical results about the dynamics of phoretic colloids close to and adsorbed at fluid interfaces in particular highlighting similarities and differences with respect to solid-fluid interfaces. 
\end{abstract}

\maketitle

\section{Introduction}\bigskip

The dynamics of active micro- and nano-sized swimmers has gained significant interest (see recent reviews such as Refs. \cite{LaugaRev,ebbens,Gompper2015_rev,Bechinger_RMP2016}) due to the large set of possible applications that spans from drug-delivery
systems~\cite{Wang_2017a,Kim2018,Halder2019,Ghosh2020} up to mimicking motile biological cells such as
bacteria~\cite{Poon}. 
In particular, an intriguing synthetic realization of micro-swimmers is the case of phoretic colloids~\cite{Paxton2004,Golestanian2005,Howse2007,Kapral2007}. Indeed, these particles attain motion in a completely different way as compared to biological swimmers. 
In fact, the latter typically attain motion by moving flagella thanks to nanometric machines, called molecular motors. In contrast, phoretic colloids move without the need of synthesizing any nanometric gear. Actually, phoretic colloids attain net displacement by inducing a net imbalance in the chemical potential of the fluid they are suspended in. 
This mechanism, firstly experimentally proposed in Ref.~\cite{Paxton2004} and then theoretically addressed in Ref.~\cite{Golestanian2005}, relies on the fact that phoretic colloids catalyze a chemical reaction on their surface. In particular, motion is attained when the rate of the chemical reaction on the surface of such particles is inhomogeneous. The performance of phoretic colloids has been addressed both theoretically~\cite{Golestanian2005,Kapral2007,
Julicher,Popescu2009,Popescu2010,Lowen2011,Seifert2012a,Koplik2013,Kapral2013,Brown2017}
as well as
experimentally~\cite{Paxton2004,SenRev,ebbens,Howse2007,Golestanian2012,Fisher2014,
Stocco}. 

Recent studies have shown that the dynamics of phoretic colloids is particularly sensitive to the presence of
boundaries and interfaces since they can affect the local imbalance in the chemical potential that eventually, for isothermal systems, is controlled by the density profile of the product of the catalytic reaction. 
Moreover, since no external force is acting on such particles, their motion is always associated (and controlled) by the motion of the fluid they move in. Accordingly, boundaries and interfaces not only affect the profile of the chemical potential but also the velocity profile of the solution and hence the motion of phoretic colloids. 
Such a feature is similar to what has been observed for micro-organisms and mechanical  swimmers~\cite{Lauga2006,Lauga2008,Lauga2014b,Gompper2015,chinappi2016}). 

\begin{figure}
 \vspace{-5pt}
  \includegraphics[scale=0.7]{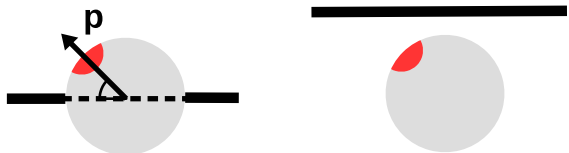}
  \caption{cartoon of an active colloid trapped at (left) or close to (right) a fluid interface (black solid line). The red spot represents the area in which the catalytic reaction takes place.}
 \label{fig:Janus-1}
\end{figure}
Here we review some recent results about the dynamics of phoretic colloids in the presence of fluid-fluid interfaces (see Fig.\ref{fig:Janus-1}). 
While both fluid-fluid and solid-fluid interfaces affect the performance of phoretic particles by deforming the chemical potential profile, fluid-fluid interfaces are characterized by some peculiar features, absent in the case of solid-fluid interfaces:\\
\textbf{Crossing.} A fluid-fluid interface can be crossed by the reaction products of the catalysis~\cite{Squarcini2020}. This is crucial because it implies that the density profile of reaction products in one fluid phase can be controlled upon tuning its dynamics in the other one. \\
\textbf{Adsorbing.} Phoretic particles can be adsorbed at a fluid-fluid interface~\cite{Stocco3}. In such a case, their collective behavior is affected by capillary interactions that are absent in the case of fluid-solid interfaces\\ 
\textbf{Responsive.} Fluid-fluid interfaces can be responsive to the imbalance in the chemical potential~\cite{Squarcini2020} and this can lead to the onset of Marangoni flows that, again, is absent in the case of solid-fluid interfaces.

The structure of the text is as follows: in Sec.~\ref{sec:therm} we set the problem of phoretic colloids in the context of linear response theory. This will allow us to highlight some common features of active systems, in general, and of phoretic colloids, in particular. Then in Sec.~\ref{sec:phoresis} we briefly review the physics of phoresis. Sections.~\ref{sec:trap},\ref{sec:close}  and Sec.~\ref{sec:marang} are dedicated to the review of the main theoretical results about the dynamics of phoretic colloids trapped at and close to fluid interfaces.
Finally, in Sec.~\ref{sec:conclusions} we derive some conclusions.

\section{Active systems from a thermodynamics perspective}\label{sec:therm}

Active systems, also called active matter, are physical systems that are in the "active" state. In order to specify the nature of this state and to identify its peculiarities it is insightful to review the diverse "states" that a system can attain.
According to thermodynamics, physical systems can be either at equilibrium or (by virtue of ``tertium non datur'') in a non-equilibrium state.
Equilibrium states are characterized by two main features~\cite{Callen_book}:
\begin{enumerate}
 \item the relevant thermodynamic variables do not change in time
 \item entropy production is identically zero
\end{enumerate}
The first requirement ensures that equilibrium state are steady states, whereas the second ensures that equilibrium steady-states do not consume energy: the system can remain in the equilibrium state forever without the need of any kind of energy or mass supply.
In contrast, many non-equilibrium states are explored by the system while relaxing towards the (possible) equilibrium state. Accordingly, these transient non-equilibrium states do not fulfill any of the two postulates. 
Interestingly, there is a particular set of non-equilibrium states, namely non-equilibrium-steady-states (NESS), which
fulfill the first postulate, however, they fail to fulfill the vanishing entropy production postulate. 
This implies that energy or mass has to be continuously supplied to keep the system in a non-equilibrium steady-state. Moreover, non-vanishing entropy production implies that the thermodynamic intensive quantities (that are pressure, temperature and chemical potentials for simple thermodynamic systems) are not homogeneous anymore and hence local fluxes will set up~\cite{DeGroot_book}.

A first attempt to tackle non-equilibrium scenarios is the linear-response/local-equilibrium approach. 
Interestingly, when the thermodynamic intensive variables vary very slowly in space one can approximate them as constant over a small region and hence, at this mesoscopic scale, the system looks like if it were in an equilibrium state characterized by the \textit{local} values of the thermodynamic intensive variables. This is the so-called local equilibrium approximation. Moreover, for mild variations of the intensive thermodynamic variables, the fluxes can be Taylor expanded and, at first order in the expansion, the fluxes are proportional to the gradients of the intensive thermodynamic variables. 
Interestingly, in such a regime (linear response regime) the transport coefficients (i.e. the prefactors relating the gradients of the intensive variables to the associated fluxes) do not depend on the forces themselves and hence can be calculated from equilibrium autocorrelation functions~\cite{DeGroot_book}.
Within this ``linear'' regime it is possible to study the dynamics of systems and therefore to characterize the non-equilibrium steady states. 
Up to now, typically systems have been driven out of equilibrium by either applying external body forces, like gravity, or by putting their boundaries in contact with reservoirs characterized by different magnitude of their temperature, pressure or chemical potential. Accordingly, in these cases entropy production is controlled by body forces and/or by the ``boundary conditions''. 

Recently, a set of systems showing a novel class of  non-equilibrium states - active systems - has been identified. The non-equilibrium state of these systems is characterized by a different origin of entropy production: entropy production is not due to external body forces or to boundary conditions, rather these systems break equilibrium locally. For example, a molecular motor hydrolyzes ATP and the energy gained from  this chemical reaction is directly transduced into mechanical work\footnote{This occurs only if the chemical potentials of ATP and ADP are not equal i.e., if the system is not at thermal equilibrium.}. It is clear that  for molecular motors motion is attained nor by an external body force neither by some boundary term, rather motion is attained due to the \textit{local} energy transduction extracted from the imbalance of the chemical potential of ATP.

Accordingly, a physical system that is at steady state can be in one of three different states:\begin{itemize}
 \item Equilibrium steady state
 \item Non-Equilibrium steady state $\begin{cases}
                                     \text{driven} \\
                                     \text{active}
                                    \end{cases}$
\end{itemize}
where with ``driven'' holds when the mechanism breaking the equilibrium is an external body force or is controlled by the boundaries whereas  ``active'' holds when the mechanism breaking the equilibrium is \textit{local}.

Clearly, due the previous definition, all living systems belong to the active class since their non-equilibrium state is driven by local imbalances in the chemical potential~\footnote{We recall that temperature gradients within living beings are, at most, rare and due to the  Gibbs-Duhem relation chemical potential gradients can be mapped into pressure gradients.}. 
However, recently also some synthetic systems capable of locally breaking  equilibrium have been designed~\cite{Ebbens_Review,Sanchez_Review,Wang_Review,Juliane_review,Fischer2018}. 
An intriguing subset of active matter is the one of micro swimmers\footnote{For an overview of other active matter systems see the recent reviews~\cite{Joanny_RMP,Sagues_review}.}: agents that can actively displace (on the top of passive diffusion). 
Typical examples spans from micrometric bacteria up to ants, fishes and mammals. A part of biological systems, a paradigmatic realization of synthetic microswimmers are phoretic colloids~\cite{Golestanian2005}. Typically, the surface of these colloids is partially  covered by a catalyst which promotes a chemical conversion of reactants (fuel)
into product molecules (See section.~\ref{sec:phoresis}). The partial coverage triggers the onset of 
density gradients along the surface which induce local fluid flows that eventually set the particle in motion. 

\begin{figure}
 \vspace{5pt}
 \centering
  \includegraphics[scale=0.55]{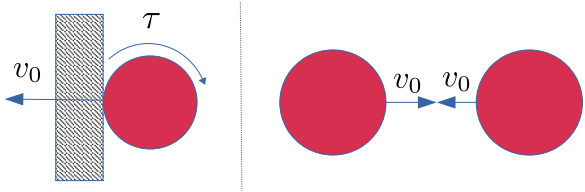}
 \caption{Schematic view of an active colloid colliding against a wall (left) or another colloid (right).}
 \label{fig:ABP}
\end{figure}

\subsection{Accumulation of micro swimmers at boundaries}
Due to their active nature, self-phoretic colloids tend to accumulate at walls and interfaces. This feature can be easily explained by the simplest theoretical model of microswimmer: the Active Brownian Particle (ABP). An ABP is a colloidal particle that undergoes both translational and rotational thermal fluctuations. On the top of the passive translational and rotational diffusion an ABP undergoes also an active displacement characterized by a velocity of magnitude $v_0$. The direction of the active velocity is determined by a unit vector whose dynamics is controlled by the rotational diffusion of the particle. Accordingly, when an ABP moves towards a wall and collides with it the ratio of the (active) force pulling towards the wall and the (thermal) force pushing away is captured by the P\'eclet number 
\begin{align}
\text{Pe}=\frac{v_0 R}{D}\,,
\end{align}
where $R$ is the particle radius and $D$ its translational diffusion coefficient. Therefore, when particles are weakly active $\text{Pe}\ll 1$ the excess of ABPs at walls is mild since the active force is overwhelmed by the thermal fluctuations. In contrast, for $\text{Pe}\gg 1$ a significant enhancement of ABPs density at walls is expected. This feature holds also when the collisions are among ABPs (see Fig.~\ref{fig:ABP}). In this case these collisions give rise to a phenomenon called Motility Induced Phase Separation (MIPS) in which ABPs separate into a gas-like phase and a solid-like phase~\cite{Cates_Review}.
These features are not peculiarities of ABPs rather they are common to all real micro swimmers. Indeed, as ABPs, real micro-swimmers are very sensitive to confining media~\cite{Xiao_Review} like boundaries~\cite{Das2015,Simmchen2016}, obstacles~\cite{Tagaki14,Zeitz17} or fluid interfaces~\cite{Fei_Review}.
In the following, we focus on the dynamics of a single confined phoretic colloid or on a dilute suspension of them.\bigskip

\section{Phoresis: a brief introduction}\label{sec:phoresis}
Phoresis (from the Greek ``to carry'')  is defined as the transport of (colloidal) particles due to the interactions with the fluid they are suspended in~\cite{Anderson1989}. 
Typically, these interactions have an electrostatic and/or Van der Waals origin and therefore their magnitude usually decays on a length scale, $\lambda$, of at most a few nanometers. Therefore, phoretic transport can be regarded as a \textit{surface} phenomenon that involves only the fluid molecules in the vicinity of the surface of the solid particle. 
According to its definition, phoretic transport involves an interplay between fluid dynamics, surface science and different kinds of transport phenomena including mass, charge and even heat flows. 
In order to clarify the onset of phoretic transport of a (spherical) colloidal particle it is insightful to zoom  in on the surface of the colloidal particle such that variations of density or velocity field on the length scale  comparable  to $\lambda$ 
can be appreciated. Due to the wide length scale separation between the typical size of a colloidal particle $\sim \mu \text{m}$ and the decay length of the surface potentials $\sim \text{nm}$ the surface of the colloidal particle can be regarded as flat. Depending on the type of surface field inducing motion diverse phoretic means can be identified: electrophoresis (controlled by electrostatic interactions with charged solutes),  diffusiophoresis (controlled by interactions with uncharged solutes), and thermophoresis (controlled by temperature gradients). In the following we derive the phoretic velocity of a colloid in the case of diffusiophoresis. Moreover, we assume that the colloid is suspended in solution characterized by an inhomogeneous density profile of solute $\rho(x,y,z)$.
\vspace*{-8pt}

\begin{figure}
\vspace*{-5pt}
 \includegraphics[scale=1.7]{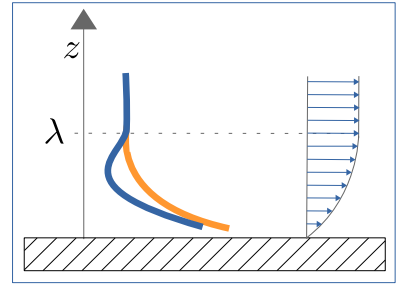}
 \caption{Schematic view of phoresis: the qualitative profiles of the  attractive (repulsive) interaction potentials between the solute molecules and the surface of the colloid (dashed area) are drawn in blue (orange) lines. We recall that, even in the case of an attractive potential, eventually the molecules experience an excluded volume interaction (i.e. repulsion) when they attempt to penetrate the surface of the colloid. The typical velocity profile is represented by arrows; we recall that the direction of the arrows depends, \textit{inter alia}, on the repulsive/attractive nature of the interaction. The typical decay length of the potential $\lambda$ is also shown. \label{fig:phoresis}}
 \vspace*{5pt}
\end{figure}\hfill%
\begin{figure*}
 \vspace{-0pt}
  \raisebox{0.135\height}{\includegraphics[scale=0.4]{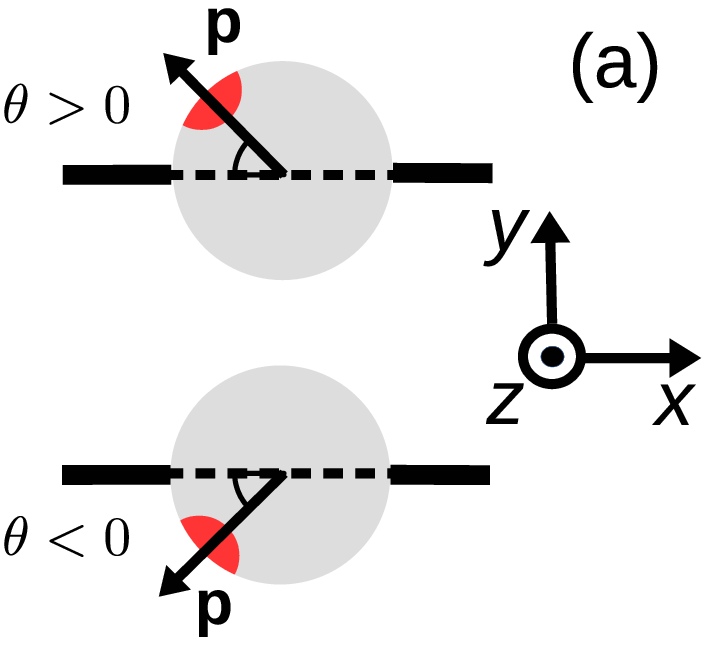}}\,\,\quad\quad
  \includegraphics[scale=0.15]{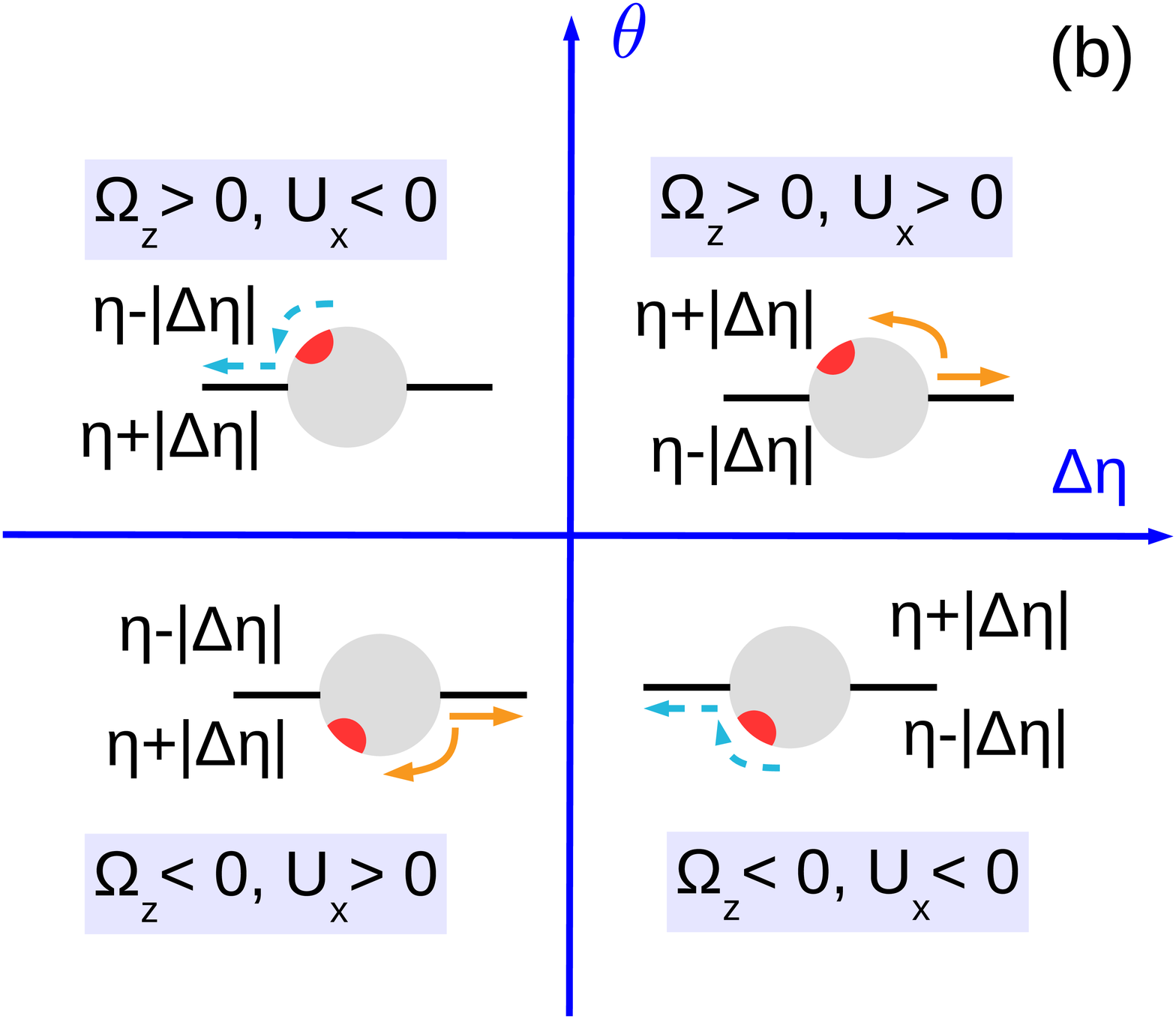}\quad
  \raisebox{0.03\height}{\includegraphics[scale=0.5]{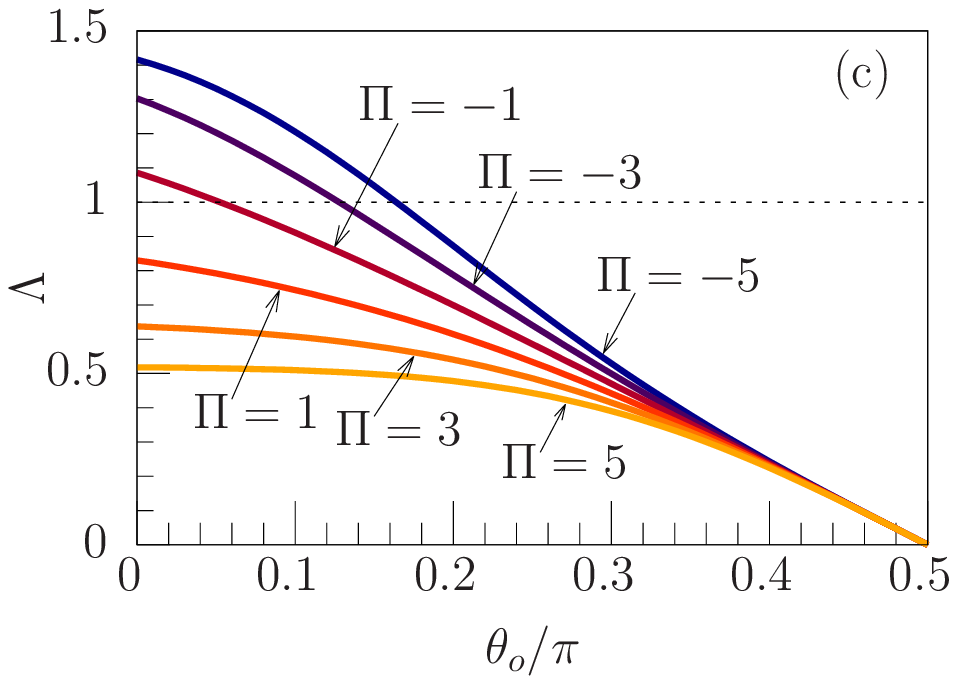}}
 \caption{(a) Definition of the orientation, $\theta$, of the Janus particle with respect to the interface.
(b) cartoon of the configurations of sustained motility as function of the viscosity difference $\Delta\eta = \eta_1-\eta_2$ between the two fluid phases and the orientation of the Janus colloid with respect to the interface, encoded in the angle $\theta$. The orange and blue colors of the arrows refer to
repulsive and attractive interactions between the chemically generated solute and the
particle, respectively. (c) ratio $\Lambda=\lambda_i/\lambda_b$
 of the persistence length of a Janus
particle at the interface ($\lambda_i$) and in the bulk
($\lambda_b$) as a function of the coverage $\theta_o$ for various values of $\Pi = \sqrt{3 \pi} \beta 
V_0 R^2 \Delta \eta$. $\theta_o$ is the opening angle of the catalytic region: for  $\theta_o=0$ the colloidal particle is fully passive, for $\theta_o=\pi/2$ the colloid is half covered (colloidal Janus particle). Figures adapted with permission from Ref.\cite{Malgaretti2016_SM}. Copyright 2016 RCS.}
 \label{fig:Janus-2}
\end{figure*}

In order to keep the notation simpler, we assume translational invariance along the direction $y$ perpendicular to the plane shown in Fig.~\ref{fig:phoresis}.
Further, we assume that the solvent is a Newtonian fluid characterized by its viscosity $\eta$. The solvent (and the solution) is assumed to be incompressible. Finally, we assume that the solute concentration varies smoothly along the longitudinal $x$ direction and that equilibrium is retained along the transverse direction $z$ (see Fig.~\ref{fig:phoresis}). When the solute is very diluted solute-solute interactions can be disregarded and its density profile, $\rho$, reads
\begin{align}
 \rho(x,z)\simeq \rho_\infty(x) e^{-\beta U(z)}
\end{align}
where $U(z)$ is the solute-wall interaction potential (see Fig.~\ref{fig:phoresis}), and $\beta=1/k_BT$ is the inverse thermal energy, being $k_B$ the Boltzmann constant and $T$ the absolute temperature.
The solute experiences a net force due to its interaction with the walls. Accordingly, the force balance requires
\vspace*{10pt}
\begin{align}
 -\frac{\partial}{\partial z}\rho(x,z)=\rho_\infty(x)\frac{\partial}{\partial z}\beta U(z)
\end{align}
and the pressure reads
\begin{align}
 \beta p(x,z) \simeq \rho_\infty(x)\left[e^{-\beta U(z)}-1\right]\,.
\end{align}
Due to the small length scales ($\sim \mu \text{m}$) and velocities ($\sim \mu \text{m/sec}$) involved, the Reynolds number, $\text{Re}=\frac{\rho v R}{\eta}$, being $R$ the radius of the colloidal particle, is small and the dynamics of the solution is captured by the Stokes equation
\begin{align}
 \eta \frac{\partial^2}{\partial z^2} v_x(x,z)=\frac{\partial}{\partial x}p(x,z)\,,
 \label{eq:stokes-here}
\end{align}
with the boundary conditions
\begin{subequations}\label{eq:stks-bc}
\begin{align}
 v_x(x,z=0)&=0\text{\,\,\,no slip on the solid surface}\\
 \frac{\partial}{\partial z} v_x(x,z)\Big |_{z\rightarrow \infty}\!\!\!\!\!\!\!\! &=0\text{\,\,\,no bulk velocity gradient}
\end{align}
\end{subequations}
The velocity profile (sketched in Fig.~\ref{fig:phoresis}) can be obtained by integrating Eq.~\eqref{eq:stokes-here} twice 
\begin{align}
 v_x(x,z)=-\frac{k_BT}{\eta}\frac{\partial}{\partial _x}\rho_\infty(x) \int\limits_0^z dz'\int\limits_{z'}^\infty \left[e^{-\beta U(z'')}-1\right]dz''\,,
\end{align}
where the integration limits are chosen in order to fulfill Eqs.~\eqref{eq:stks-bc}. Since the potential $U(z)$ is short-ranged with range $\lambda$ the last expression can be approximated leading to the phoretic slip velocity
\begin{align}
 v_x(x)=-\frac{k_BT}{\eta}\mathcal{L}\frac{\partial}{\partial _x}\rho_\infty(x)\,,
\end{align}
where we have identified the so-called phoretic mobility
\begin{align}
  \mathcal{L}=\int_0^z dz'\int_{z'}^\infty \left[e^{-\beta U(z'')}-1\right]dz''\,.
\end{align}
\textbf{Self-phoretic (active) colloids}\\
In the above derivation the colloidal particle moves due to the density profile $\rho(x,z)$ that is induced by some mechanism external to the colloidal particle. However, it is possible to functionalize the surface of a  colloidal particle such that it can catalyze some chemical reaction that exploits the suspended reactants. For partial covering of the colloidal surface, the  reaction rate of the chemical reaction is inhomogeneous and hence it induce an inhomogeneous density profile of both reactants and reaction products. Accordingly, these colloids are called self-phoretic (or active) colloids since they induce the density gradient that induces their own motion~\cite{Golestanian2005}. 

Several theoretical~\cite{Golestanian2005,Kapral2007,
Julicher,Popescu2009,Popescu2010,Lowen2011,Seifert2012a,Koplik2013,Kapral2013,Brown2017}
and experimental~\cite{Paxton2004,SenRev,ebbens,Howse2007,Golestanian2012,Fisher2014,
Stocco,Stocco2021} studies have characterized the performance of self-phoretic colloidal particles. 
In particular, the active displacement of self-phoretic colloidal particles is controlled by the local gradients around the particle and therefore is quite sensitive to the presence of boundaries, obstacles or fluid interfaces in the vicinity of the self-phoretic colloidal particle. For example, it has been shown that the presence of boundaries can modulate the net velocity~\cite{Popescu2009,Yariv2016,Chang2016} and wall-bounded steady states can be induced~\cite{Uspal2015,Yariv2016,Ibrahim2016,Sharifi-Mood2016}. These features have been exploited to guide the motion of self-phoretic colloidal particles 
\cite{Das2015,Simmchen2016,Uspal2016}. Similar effective interactions have appeared for interactions among self-phoretic colloidal particles~\cite{Popescu2010,Michelin2015,Koplik2016b,Parvin2016}. Recently, experimental~\cite{Stocco,Isa2017,Stocco2017} and theoretical~\cite{Masoud2014,Wurger2014,Stark2014,Malgaretti2016_SM,Dominguez2016,Dominguez2016_2,Malgaretti2018_SM} 
studies have started to tackle the issue of motion of self-phoretic colloidal particles near or trapped at a liquid-fluid interface.


\section{Phoretic colloids trapped at fluid interfaces}\label{sec:trap}
\begin{figure}
 \vspace{5pt}
 \centering
  \raisebox{0.03\height}{\includegraphics[scale=0.8]{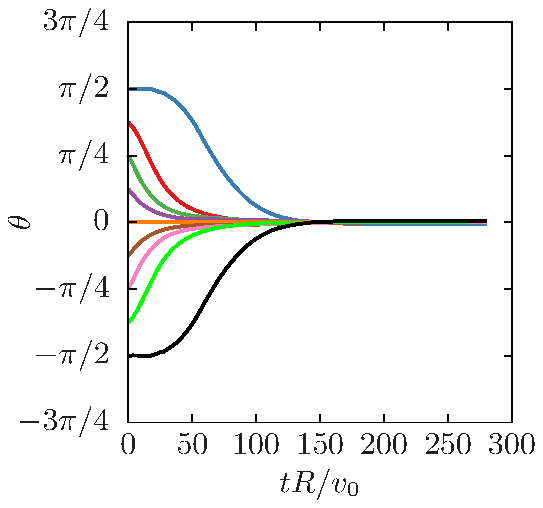}}
  \includegraphics[scale=0.8]{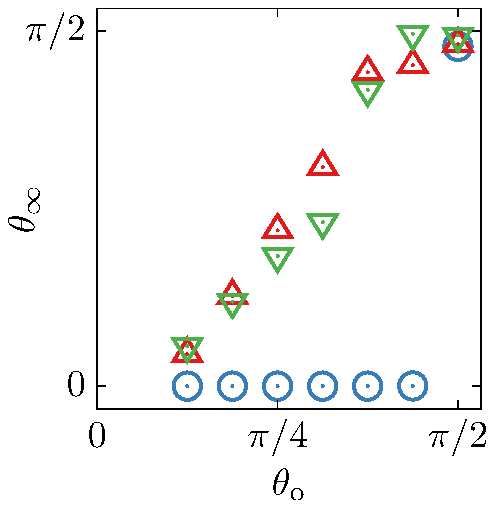}
  \caption{Left: time evolution of the orientation, $\theta$, of a catalytic colloid trapped at a fluid interface in the case in which both fluid phases have the same viscosity. Right: steady state orientation, $\theta_\infty$, as function of the covering angle $\theta_o$ and for diverse contact angles, $\theta_a=0.5\pi, 0.55\pi, 0.6\pi$ for blue circles, red upward triangles, and green downward triangles,  respectively. 
  Figures adapted with permission from Ref.~\cite{Malgaretti2020}. Copyright 2020 RCS.}
 \label{fig:Janus-22}
\end{figure}
In thermal equilibrium (i.e., in the absence of diffusiophoresis), a 
Janus particle trapped at a fluid interface typically exhibits a configuration in which the 
particle axis is not aligned with the normal of the interface (see Fig. \ref{fig:Janus-1} and Ref.~\cite{Stocco3} for experimental results). 
Therefore, upon turning on the chemical reaction, motion along the 
interface may be achieved. However, the motion at a fluid  
interface generally involves a coupling between translation and 
rotation \cite{LaugaReview,Brenner_Book,Pozrikidis2007}. Thus, the possibility arises that the 
translation along the interface may lead to a rotation of the axis of the particle 
towards alignment with the interface normal.
In Ref.\cite{Malgaretti2016_SM}, authors put forward an approximated analytical model in which the ``activity'' of the colloidal particle is accounted for by the phoretic slip velocity (see Sec.~\ref{sec:phoresis}). 
This model provides insight into the conditions under which translation occurs along the interface.

\noindent Interestingly, Fig.~\ref{fig:Janus-2} shows that a motile state, i.e. sliding along the interface, can be promoted by controlling the viscosity contrast between the two fluid phases~\cite{Malgaretti2016_SM}. This is crucial for experimental set ups where it is hard to modify the properties of the fluid in which the ``fuel'' molecules are catalyzed without changing the catalytic rate. From this perspective, these results show that the motile state of the particle can be controlled by tuning the \textit{other} fluid phase in which no catalysis takes place. 

Once steady motion is attained, the next question is about its magnitude. Are catalytic colloids trapped at fluid interfaces moving faster or slower than in the bulk? In this regard an interesting observable is the persistence length
\begin{align}
  \lambda_P=v\tau
\end{align}
where $v$ is the velocity of the colloid at the interface and $\tau$ is the rotational diffusion time. 
The persistence length captures, in the long time limit, the departure from equilibrium diffusion due to activity, since passive colloidal particles are characterized by $\lambda=0$~\cite{Stark2016}. 
Interestingly, Fig.~\ref{fig:Janus-2}c shows that the persistence length of active colloids trapped at fluid interfaces can be enhanced as compared to the persistence length of the same particle in homogeneous unbound fluids.

Typically, the description of the dynamics of active colloids is performed via a standard coarse-grained approach, within which the relative velocity between the particle and the fluid is accounted for by the so-called phoretic slip velocity (see Sec.~\ref{sec:phoresis}) on the surface of the particle~\cite{Anderson1989,Golestanian2005,Popescu2009,Julicher2009,Poon2013}. However, when an active colloid is adsorbed at a fluid interface such an approach becomes more complicated because of the presence of a three-phase contact line. 
In Ref.\cite{Malgaretti2020}, authors present  a novel numerical approach within  which the motion of the self-diffusiophoretic colloid is obtained by using the lattice Boltzmann method~\cite{Benzi1992,Krueger_book,Harting2016}, in order to construct approximate solutions of the Navier-Stokes equation directly. Within the scheme of Ref.\cite{Malgaretti2020}, this hydrodynamics solver is combined with an advection and diffusion equation for the reactants.
Such an approach allows one to discuss the reliability of the slip-velocity approach by comparison with the approximate analytical results of Ref.\cite{Malgaretti2016_SM}.
Indeed, the results of Ref.\cite{Malgaretti2020} show that, self-phoretic colloids trapped at fluid interfaces reorient their symmetry axis even in the case of equal viscosity fluids, see Fig.~\ref{fig:Janus-22}. This reorientation occurs whenever the axis of symmetry of the particle is not perpendicular or parallel to the interface. Indeed, for these cases the presence of the interface affects the velocity profile and leads to net torques on the particle. 
\begin{figure*}
 \vspace{5pt}
 \includegraphics[scale=.8]{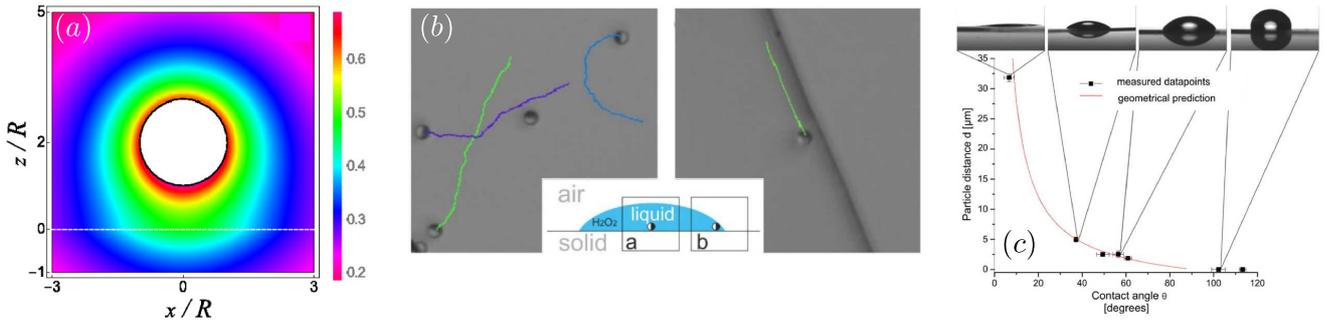}
 \caption{(a): cartoon of the density profile induced by an active colloid fully covered by catalyst close to the interface between two fluid phases characterized by different diffusion coefficients of the reaction product $D(z>0)<D(z<0)$. Figure adapted with permission from Ref.~\cite{Malgaretti2018_SM}. Copyright APS 2016. (b): top view micrograph indicating the motion of active particles on a non-treated glass surface within the drop (a) and at the edge of the drop (b), the inset represents a schematic explaining the particle position within the drop. (c): contact angle extracted from the position of the particle (points) and the prediction of the analytical model (line). Figures adapted with permission from Ref.~\cite{Simmchen2017}. Copyright 2020 Wiley.}
 \label{fig:Janus-3}
\end{figure*}
\section{Phoretic colloids close to fluid interfaces}\label{sec:close}

In many circumstances active colloids are not (yet) absorbed at fluid interfaces. However, their dynamics can be affected by the fluid interface when active colloids are nearby. In order to highlight the effect of the fluid interface on the dynamics of nearby active colloids it is insightful to study the case of a colloidal particle which is homogeneously covered by catalyst in the vicinity of a fluid interface~\cite{Malgaretti2018_SM}. Due to its spherical symmetry, the particle releases the products of the catalytic reaction isotropically and hence no net velocity is expected in a homogeneous and unbound fluid (see left panel of Fig.~\ref{fig:Janus-3}). Therefore any motion is caused by the effective interaction with the interface. Indeed, the presence of an interface breaks the homogeneity of the diffusivities of the reaction products in the two fluid phases which leads to an inhomogeneous distribution of the concentration of the reaction products along the interface normal.
Accordingly, such an inhomogeneous density profile leads to an interface-induced phoresis, the direction of which is normal to the interface, similarly to what has already been reported for the case of a hard wall~\cite{Uspal2015,Yariv2016}. Interestingly, in the present case the sign of the resulting velocity depends not only on the surface properties of the particle, as it is the case near a hard wall, rather it depends on both the contrast between the diffusivities and the distinct solubility of the catalysis products in the two fluid phases. Such predictions can, possibly, be verified using setups similar to that discussed in Refs.\cite{Jalilvand2020,Simmchen2021}.

As highlighted in Sec.~\ref{sec:phoresis}, the motion of active Janus particle leads them to accumulate at boundaries (a situation similar to the case depicted in the right panel of Fig.\ref{fig:Janus-1}). This feature can be exploited to detect and characterize such edges. For example, when active Janus particles are suspended within a droplet sitting on a solid substrate  (see central and right panel of Fig.~\ref{fig:Janus-3}) they accumulate in the vicinity of the three phase contact line. Since getting absorbed or even crossing the fluid interface requires to overcome a free energy barrier, active colloids typically remain in the liquid phase and move along the edges of the droplet. Interestingly, such a feature has been exploited to measure the contact angle of sessile droplets. Indeed, in Ref.~\cite{Simmchen2017}, authors determine the \textit{local} contact angle of the  droplet by measuring their distance from the edge of the droplet.

\section{Phoretic colloids and reactive interfaces}\label{sec:marang}

An intriguing and peculiar feature of liquid-liquid interfaces is that they can be much more responsive to stimuli than liquid-solid interfaces. 
In fact, fluid interfaces are constituted by molecules in the liquid state and hence they are highly mobile. However, the  Hamiltonian interactions that keep the two phases separate are typically quite strong as compared to thermal energy and hence fluid interfaces typically look very still. The strength of the effective interactions that keep the phases separated is encoded in the \textit{surface tension}.
However, when the strength of these interactions is (locally) modified also the physical properties of the interface  change.
One of the most known (and appreciated by sommeliers) response of fluid interfaces to inhomogeneous stimuli is the so-called Marangoni flow. Indeed, inhomogeneities of the local densities of the two fluid phases induce an inhomogeneous surface tension that eventually, by inducing local stresses on the fluid phases, leads to the onset Marangoni flow~\cite{Landau_book}. 
Such modulations of the density may be caused by the inhomogeneous density profile of reaction products induced by a phoretic colloid (see Sec.~\ref{sec:phoresis}). 
\begin{figure}
 \vspace{5pt}
 \centering
   \includegraphics[scale=0.8]{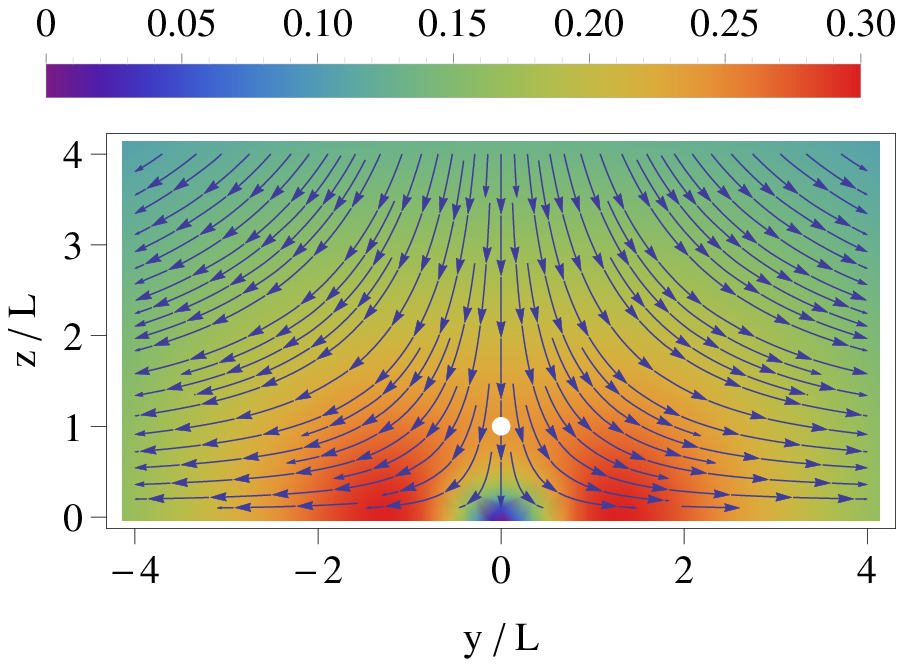}
  \includegraphics[scale=0.4]{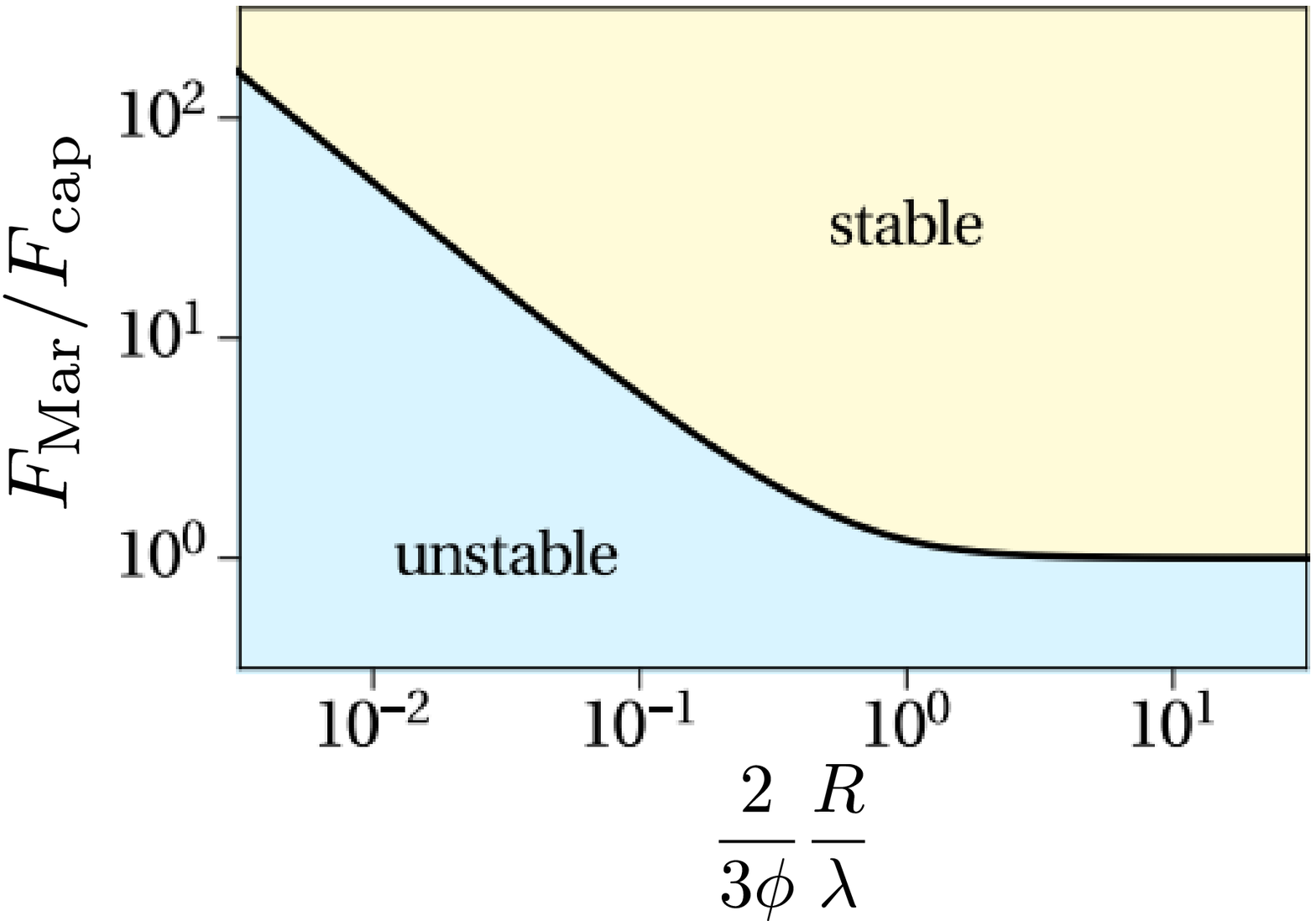}
  \caption{Top: density profile (color-coded) induced by a homogeneous catalytic colloidal particle (white dot) close to a fluid-fluid interface (located at $z=0$). Arrows represent the velocity profile (Marangoni flow) in the upper fluid phase induced by local imbalance in the surface tension at the interface. Figure reproduced with permission from Ref.~\cite{Dominguez2016}. Copyright APS 2016.
  Bottom: stability diagram in the parameter space spanned by the ratios $F_\text{mar}/F_\text{cap}$ and $\frac{2}{3\phi}\frac{R}{\lambda}$ where $\lambda=\sqrt{\gamma/\rho g}$ is the capillary length, $\gamma$ the surface tension and $\rho g$ the gravitational force density. The ''stable'' region marks the set of parameters for which the Marangoni flows stabilize the monolayer, whereas the opposite holds for the ``unstable'' region. Figure adapted with permission from Ref.~\cite{Dominguez2016_2}. Copyright RCS 2016.
  }
 \label{fig:Janus-4}
\end{figure}
Interestingly, as far as the Marangoni flow is operational (i.e. when the solute density significantly affects the surface tension), the interface-induced phoresis (see Sec.\ref{sec:close}) is subdominant whereas for those interfaces that are not sensitive to the solute density the interface-induced phoresis is the dominant effect. 
We remark that even though for both, interface induced phoresis and Marangoni flow the net motion of the phoretic colloids is along the normal at the interface, the physical mechanism underlying these two phenomena is completely different. In fact, in the case of Marangoni flow the net motion is due to the imbalance of the stress at the fluid interface (induced by the inhomogeneous surface tension),  whereas for the interface induced phoresis the motion is due to a phoretic process triggered at the surface of the colloid and induced by the asymmetry of the solute density about the interface.
In particular, both far-field approximation and exact solutions show that such an effect is prominent for liquid-liquid interfaces (such as water-oil) whereas it is negligible for liquid-gas interfaces (such as water-air)~\cite{Dominguez2016}. Remarkably, such a feature has been recently proven experimentally for both macroscopic surfers~\cite{Sur2019} as well as for micrometric droplets~\cite{Singh2020} and flat interfaces~\cite{Wittmann2021}.

This interplay between active colloids and fluid interfaces naturally raises the issue concerning the collective, i.e., large-scale behavior of a grand collection of chemically active particles forming a monolayer at a fluid-fluid interface.  The effective interaction between two particles, a distance $d$ apart, due to the advection by the induced Marangoni flow is equivalent to a long-ranged interparticle force decaying asymptotically as $\sim 1/r$. 
Such a long-range interaction can indeed compete with the other typical interaction between colloid trapped at a fluid interface: 
capillary interactions~\footnote{In principle, also the effective interactions that the inhomogeneous solute profile induces on the colloids (phoretic interactions~\cite{Liebchen2019,Scagliarini2020}) should be accounted for. However, in the presence of Marangoni flows such interactions are expected to be subdominant. This argument follows from the comparison of the Marangoni-induced interaction between a phoretic colloid and a fluid interface (Ref.~\cite{Dominguez2016}) as compared to the case of interface-induced phoresis (Ref.~\cite{Malgaretti2018_SM}).}.
Interestingly, via far-field approximations, it is possible to show~\cite{Dominguez2016_2} that when the effective force induced by the 
Marangoni flow, $F_{Mar}$ is much larger than the capillary force, $F_{cap}$, the colloidal suspension at the interface is stabilized and the clustering instability driven by the capillary attraction is prevented  (see Fig.\ref{fig:Janus-4}).  
Accordingly, such a competition between $F_{Mar}$ and $F_{cap}$ can be exploited to tune self-assembly at fluid interfaces.

\section{Conclusions}\label{sec:conclusions}

In this contribution we have reviewed the dynamics of phoretic colloids in the presence of fluid interfaces. 
In particular, we have analyzed the two possible scenarios: the case in which phoretic colloids get adsorbed at the interface and the case in which they move in the proximity of the interface. Moreover, we have also accounted for the case in which the fluid interface is responsive to the inhomogeneous density profile induced by the phoretic colloidal particles. 

In the case of phoretic colloids \textit{adsorbed} at fluid interfaces we have discussed the dependence of their orientation with respect to the interface upon varying the viscosity contrast, $\Delta \eta$ between the fluid phases. For $\Delta \eta \neq 0$, linear displacement gets coupled to angular reorientation.  Interestingly, this paves the route for a possible control of the motion of active colloids upon tuning the viscosity of the phase in which no catalysis occurs i.e., without affecting the catalytic process. 
At variance, for $\Delta \eta=0$ the simplified analytical model predict no roto-translational coupling. 
Such results have been checked against numerical simulations. Interestingly, numerical results show that also for $\Delta \eta=0$ rotation and translation are coupled via to the boundary condition that the interface imposes on the solution of the Stokes equation. 

In the case of phoretic colloids \textit{close} to fluid interfaces we have reviewed diverse scenarios. Firstly, we have remarked that the presence of the interface breaks the fore-aft symmetry along the normal to the interface that induces an inhomogeneity in the diffusion coefficient of the reaction products of the catalysis. This leads to an asymmetric density profile that can affect the dynamics of Janus colloids and it may even set isotropic phoretic colloids on motion. 
This accumulation of phoretic colloids at fluid interfaces has been already exploited to characterize the local properties, such as the local contact angle of fluid interfaces.  

Finally, we have addressed the case in which the fluid interface is sensitive to the local density profile of reaction products of the catalysis. In such a case, strong Marangoni flow will set up hence leading to an effective attraction (repulsion) of the phoretic colloid to (from) the interface. On the top of this, when several colloids are close to a fluid interface, the Marangoni flows give raise to effective interactions among the phoretic colloids. Interestingly, these effective interactions are long-ranged, $\propto 1/r$, and hence compete with capillary interactions. In the case of colloids adsorbed at fluid interfaces, such Marangoni-induced interactions among colloids can prevent the capillary collapse of colloidal suspensions. 

\section*{Acknowledgments}

We acknowledge funding by the Deutsche Forschungsgemeinschaft (DFG, German Research Foundation) – Project-ID 416229255 – SFB 1411. 

\section*{Keywords}

Active Matter, Fluid Interfaces, Microswimmers, Phoretic colloids.

\bibliography{juliane_biblio}
\end{document}